\documentclass[letterpaper]{article} 
\usepackage{aaai25}  
\usepackage{times}  
\usepackage{multirow} 
\pdfoutput=1
\usepackage{booktabs}
\usepackage{bbding}
\usepackage{amssymb}
\usepackage{helvet}  
\usepackage{courier}  
\usepackage[hyphens]{url}  
\usepackage{graphicx} 
\usepackage{natbib}  
\usepackage{caption} 
\usepackage{algorithm}
\usepackage{algorithmic}

\usepackage{newfloat}
\usepackage{listings}
\DeclareCaptionStyle{ruled}{labelfont=normalfont,labelsep=colon,strut=off} 
\floatstyle{ruled}
\newfloat{listing}{tb}{lst}{}
\floatname{listing}{Listing}
\title{BSDB-Net: Band-Split Dual-Branch Network with Selective State Spaces Mechanism for Monaural Speech Enhancement\\}
\author{
    Cunhang Fan\textsuperscript{\rm 1},
    Enrui Liu\textsuperscript{\rm 1},
    Andong Li\textsuperscript{\rm 2}$^,$\textsuperscript{\rm 3}\equalcontrib,
    Jianhua Tao\textsuperscript{\rm 4},
    Jian Zhou\textsuperscript{\rm 1},
    Jiahao Li\textsuperscript{\rm 1},
    Chengshi Zheng\textsuperscript{\rm 2}$^,$\textsuperscript{\rm 3},
    Zhao Lv\textsuperscript{\rm 1}\equalcontrib
}
\affiliations{
    \textsuperscript{\rm 1}Anhui Province Key Laboratory of Multimodal Cognitive Computation, \\School of Computer Science and Technology, Anhui University, Hefei, China\\
    \textsuperscript{\rm 2}Key Laboratory of Noise and Vibration Research, Institute of Acoustics Chinese Academy of Sciences, Beijing, China\\
    \textsuperscript{\rm 3}University of Chinese Academy of Sciences, Beijing, China\\ 
    \textsuperscript{\rm 4}Department of Automation, Tsinghua University, Beijing, China\\

    cunhang.fan@ahu.edu.cn, e23201090@stu.ahu.edu.cn, liandong@mail.ioa.ac.cn, jhtao@nlpr.ia.ac.cn, \\jzhou@ahu.edu.cn, e23301209@stu.ahu.edu.cn, cszheng@mail.ioa.ac.cn, kjlz@ahu.edu.cn
    
}

\usepackage{bibentry}

\begin{document}

\maketitle

\begin{abstract}
Although the complex spectrum-based speech enhancement (SE) methods have achieved significant performance, coupling amplitude and phase can lead to a compensation effect, where amplitude information is sacrificed to compensate for the phase that is harmful to SE. In addition, to further improve the performance of SE, many modules are stacked onto SE, resulting in increased model complexity that limits the application of SE. To address these problems, we proposed a dual-path network based on compressed frequency using Mamba. First, we extract amplitude and phase information through parallel dual branches. This approach leverages structured complex spectra to implicitly capture phase information and solves the compensation effect by decoupling amplitude and phase, and the network incorporates an interaction module to suppress unnecessary parts and recover missing components from the other branch. Second, to reduce network complexity, the network introduces a band-split strategy to compress the frequency dimension. To further reduce complexity while maintaining good performance, we designed a Mamba-based module that models the time and frequency dimensions under linear complexity. Finally, compared to baselines, our model achieves an average 8.3 times reduction in computational complexity while maintaining superior performance. Furthermore, it achieves a 25 times reduction in complexity compared to transformer-based models.
\end{abstract}

%

\section{Introduction}
In the realm of audio signal processing, speech enhancement (SE) is regarded as a fundamental technique to recover the clean speech from noisy environments. The degradation of speech quality by background noise is not only perceptually bothersome but also significantly impairs the performance of automatic speech recognition (ASR)~\cite{c:1, c:2}. Also, SE is indispensable in smart devices, vehicular systems, and home automation. With the burgeoning prevalence of online conferencing, the demand for real-time SE solutions has surged, underscoring the necessity for techniques that are both effective and computationally efficient~\cite{c:3}.

Existing speech enhancement methods can be roughly categorized into two classes, namely in the time domain~\cite{r:4, c:5} and in the time-frequency (T-F) domain. This paper mainly focuses on the latter. Initially, SE methods focused on magnitude spectrum enhancement with the phase kept unaltered~\cite{GCRN}. Subsequent researches have revealed the pivotal role of phase~\cite{c:5}, which inspire many studies to model complex spectra to better recover phase information. Masking-based methods, such as the complex ratio mask~\cite{CRM}, have been recognized for their ability to modulate both the real (R) and imaginary (I) components of the noisy complex spectra, surpassing the performance of traditional masks like ideal binary mask~\cite{ibm} and ideal ratio mask~\cite{IRM}. Besides, complex spectral mapping (CSM) has also been introduced to reconstruct the target RI components directly~\cite{GCRN}. However, this direct mapping can lead to a compensation effect between phase and magnitude~{\cite{wang2021compensation}}. Recent advancements have seen the application of decoupling concepts in audio processing~\cite{DTF-AT}. Some multi-stage methods have been proposed to address SE problems. These methods decouple the original mapping problem into two stages: predicting the magnitude spectrum at first and refining the complex spectrum through residual learning in the second stage. This approach partially mitigates the implicit compensation effect between phase and magnitude~\cite{ctsnet}. Nonetheless, the sequential nature of these methods can be limiting, as the performance of later stages is heavily dependent on the output of earlier stages~\cite{gagnet}.

Despite the promising results of previous works, their high computational complexity can impede the practical application of SE models. First, as a front-end task, the impact of SE on downstream tasks, such as ASR, should be considered. Moreover, effective deployment in edge or resource-constrained environments, like online meetings and real-time communications, necessitates extremely low-complexity SE methods. Strategies such as frequency band division~\cite{sprnn}, have been introduced to reduce frequency modeling overhead. Additionally, the computational complexity of sequence modeling can be substantial, with some transformer-based models reaching hundreds of gigaflops (G/s) due to their quadratic attention mechanisms~\cite{MP-SENetconformer}. Recently, State Space Models~\cite{ssm}, such as HIPPO~\cite{gu2020hippo} and the Structured State Space Model~\cite{s4}, have demonstrated promising performance with reduced complexity. The Selective SSM~\cite{gu2023mamba}, in particular, enables the establishment of long-range dependencies with linear computational complexity, making it suitable for sequence modeling.

In light of these challenges, we propose the Band-Split Dual-branch Network (BSDB-Net) for the monaural SE task. This approach introduces a decomposition strategy to enhance the magnitude and complex spectra in parallel. Specifically, we devised the Magnitude Enhancement Network (MEN) to suppress the noise components without interfering with phase information, and the Complex Spectral Enhancement Network (CEN) to restore phase information implicitly, complementing the magnitude features extracted by MEN and mitigating the compensation effect. Besides, to effective decrease the overhead for frequency modeling, a frequency band-split strategy is adopted to compress the frequency axis. To better decrease the complexity for sequence modeling, the Mamba structure is introduced to reduce complexity in temporal and frequency sequential relations.

The major contributions of this paper are summarized as follows:
\begin{itemize}
    \item We propose a dual-branch parallel speech enhancement network that implicitly extracts phase information while ensuring the independence of amplitude information.
    \item We adopt a frequency band-splitting strategy and Mamba-based sequence modeling modules, significantly reducing computational complexity.
    \item Comprehensive experiments on two public datasets demonstrate that BSDB-Net achieves an average \textbf{8.3} times reduction in computational complexity while maintaining comparable performance.
\end{itemize}

\begin{figure*}[t]
\centering
\includegraphics[width=1\textwidth]{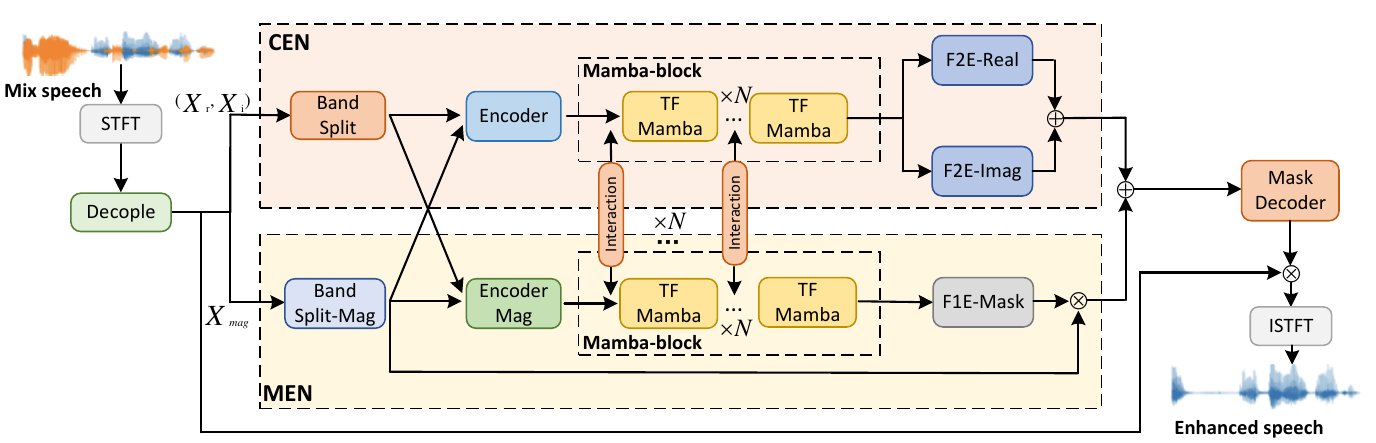} 
\caption{Overall architecture of the proposed BSDB-Net consists of three main components. The first part includes the Band-Split module for frequency band segmentation and the Mask-Decoder module for generating masks used in band synthesis. The second part features a dual-branch enhancement network: the MEN branch suppresses noise in the magnitude spectrum roughly, while the CEN branch primarily estimates complex spectra to capture phase characteristics. The third part involves the Mamba-block module designed for sequence modeling.}
\label{fig:fig1}
\end{figure*}

\section{Related Work}
\textbf{Multi-stage Methods}: Due to the lack of prior information, the performance of the single-stage SE pipeline is often heavily limited in complicated acoustic scenarios. In contrast, for the multi-stage pipeline, the original mapping problem is usually decomposed into several separate subtasks to enable the learning progressively~\cite{li2022taylor}. In DTLN~\cite{lstmdualsignall}, the authors proposed a stacked dual signal transformation network. In FullSubNet~\cite{fullsubnet}, the idea of combing sub-band and full-band was proposed to restore the spectrum. CompNet~\cite{compnet} combined the time domain and T-F domain, using a cross-domain complementary approach to optimize the speech enhancement network. CTS-Net~\cite{ctsnet} utilized a two-stage paradigm to supplement phase information on the basis of extracting the magnitude spectrum. TaylorNet~\cite{li2022taylor} proposed an end-to-end framework to simulate the 0th-order and high-order items of Taylor-unfolding. FAF-Net~\cite{FAFNET} proposed reference-based speech enhancement via a feature alignment and fusion network.

\textbf{Sequence Modeling}: The exploration of long and short-term temporal dependencies within speech signals prompted the adoption of Recurrent Neural Networks (RNNs~\cite{rnn}) to better capture contextual relations. To address the issue of gradient explosion, Long Short-Term Memory (LSTM~\cite{LSTM}) networks were introduced. The temporal convolutional module (TCM)~\cite{TCM} introduced in TCM was found to be more effective in time series modeling than LSTM. Recently, a compressed version of TCM called the squeezed temporal convolutional module (S-TCM)~\cite{ctsnet} was proposed. Transformer-based models~\cite{Selector-enhancer} have emerged as a promising alternative due to their superior capability in modeling long-range dependencies using self-attention mechanisms~\cite{DBTNET}. Due to its linear complexity, Mamba~\cite{li2024spmamba} is considered a promising alternative to transformer in sequence modeling.

\section{Proposed Architecture}
The received noisy speech in the short-time Fourier transform (STFT) domain can be presented as,
\begin{equation}
Y(t,f)=X(t,f)+N(t,f)
\end{equation}
where $\left\{Y, X, N\right\}$ denote the mixture, clean and noise, respectively. $t\in\left\{1,\cdots,T\right\}$ is the time frame index, and $f\in\left\{1,\cdots,F\right\}$ is the frequency index.

The proposed BSDS framework is shown in Figure~{\ref{fig:fig1}}. The noisy magnitude and complex spectra are first decoupled. By separately band-splitting and feature encoding, they are converted into abstract representations for magnitude and phase, respectively. Then, stacked T-F Mamba-blocks are adopted for effectively modeling along the time and frequency axes. Subsequently, the representations from two streams are fused. The segmented frequency bands are then merged using a Mask-Decoder module to obtain the estimated complex spectrum. 

\subsection{Band-Split and Mask-Decoder}
\begin{figure}[t]
\centering
\includegraphics[width=1\columnwidth]{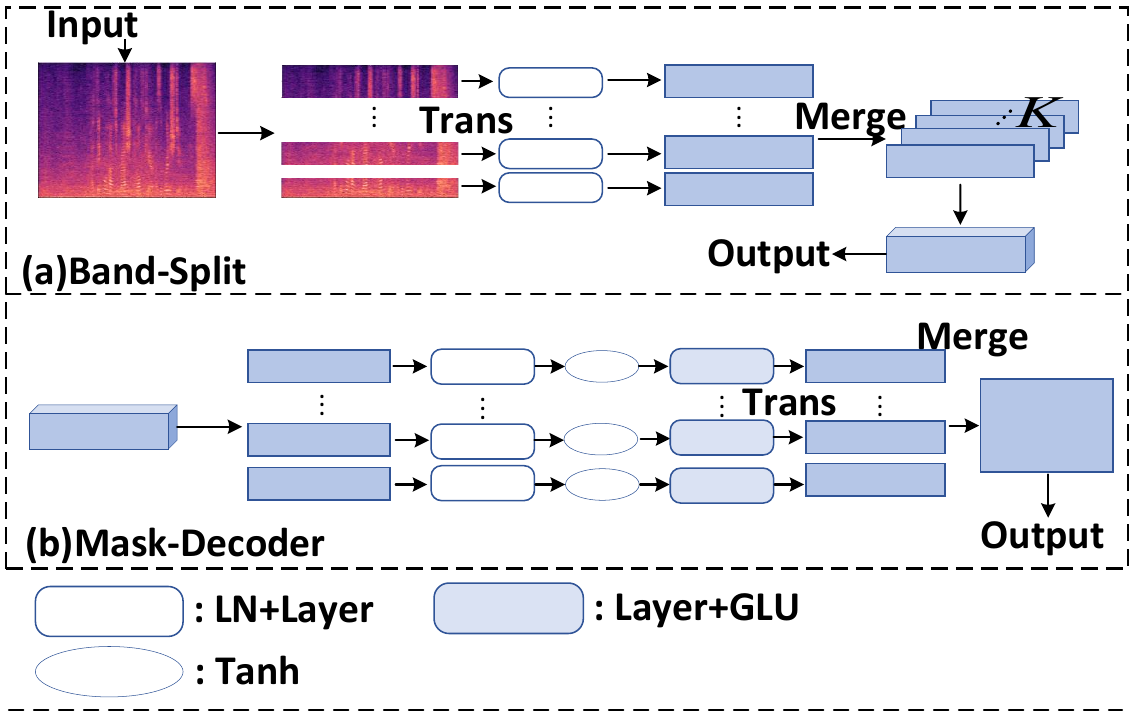} 
\caption{(a) The Band-Split module divides frequency bands for input into the modeling module. (b) The Mask-Decoder module synthesizes frequency bands post-modeling to generate masks.}
\label{fig:fig2}
\end{figure}

As shown in Figure~{\ref{fig:fig2}}, the complex and magnitude spectra are first compressed into lower-resolution bands using the Band-Split module~{\cite{sprnn}}. To be specific, the noisy input spectrum $X$ is segmented into a sequence of non-overlapping frequency bands $\left \{ A_{i} \right \}_{i=1}^{K}$, each of which is individually projected to yield the embedding of dimension $N$. Subsequently, the $K$ bands are stacked to obtain a 3-D tensor $Z$. The process can be formulated into
\begin{equation}
A_{i}\in \left \{ A_{1}, A_{2},..., A_{K} \right \}, A_{i}\in \mathbb{R}^{T\times F_{i}}
\end{equation}
\begin{equation}
X_{i}=FC(LN(A_{i})), X_{i}\in \mathbb{R}^{T\times N}
\end{equation}
\begin{equation}
	Z = Concat\left(X_{1}, X_{2},..., X_{K}\right)\in \mathbb{R}^{K\times T\times N}
\end{equation}
where $A_{i}$ represents the $i$ frequency bands to be split, $\left\{FC, LN\right\}$ denote the linear layer and layer normalization, respectively. $Concat\left(\cdot\right)$ denotes the concatenation operation.

For the Band-Merge module, let us denote the input as $B\in\mathbb{R}^{K\times T\times N}$. 
Similarly, for the frequency band feature $B_{i}\in\mathbb{R}^{T\times N}$, where $i\in\left\{1,\cdots,K\right\}$ denotes the band index, it is processed through separate layers to obtain the decoded target feature. After all the bands are processed, they are concatenated along the frequency axis to form the output $M$, whose formulation can be given by
\begin{equation}
M_{i} = GLU(Tanh(FC(LN(B_{i}))))
\end{equation}
\begin{equation}
M = Merge({M_{1}, M_{2},..., M_{K}}), M\in \mathbb{R}^{F \times T}
\end{equation}
where $B_{i}$ denotes the $i$th band feature, $\left\{Tanh, GLU\right\}$ denote the Tanh activation function and gated linear unit, respectively. $Merge\left(\cdot\right)$ is the concatenation operation along the frequency axis. 

\subsection{Dual-Branch}
After the band split module, the magnitude-oriented and complex-oriented features are fed into the MEN and CEN branches for modeling, respectively.

For both branches, the data first pass through the encoder layer. Instead of using typical encoder layers with down-sampling operations, here we coarsely model the features without frequency downsampling to mitigate the possible information loss:
\begin{equation}
C_{m}=PReLU(LN(Conv2d(Inter(Z_{mag}, Z_{ri}))))
\end{equation}
\begin{equation}
C_{ri}=PReLU(LN(Conv2d(Inter(Z_{ri}, Z_{mag}))))
\end{equation}
where $Z_{mag}$, $Z_{ri}$ represent the magnitude and complex features after Band-Split module, $C_{m}$, $C_{ri}$ denote the output of magnitude and complex encoders, respectively. $Inter\left(a, b\right)$ denotes the interaction module between the features $a$ and $b$.

As shown in Figure~{\ref{fig:fig1}}, the feature processed by the encoder is passed to the sequence modeling using the Mamba-Block. After that, a mask is generated via the F1E-Mask layer. The F1E Mask refers to the extraction of features that include amplitude and the generation of a mask. It includes a dilated block layer to expand the model's receptive field, followed by the PReLU activation function and layer normalization. Finally, the input data to the MEN and the estimated mask are multiplied to obtain the filtered feature of the MEN branch. The whole process can be given by
\begin{equation}
D= PReLU(LN(DConv(Mam(C_{mag}))))
\end{equation}
\begin{equation}
D_{1}=Sigmoid(Conv(D))
\end{equation}
\begin{equation}
D_{2}=Tanh(Conv(D))
\end{equation}
\begin{equation}
D_{meg} = D_{1} \otimes D_{2}, MEN_{out} = A_{mag} \otimes D_{meg}
\end{equation}
where $\otimes$ denotes the elementary multiplication operation. $\left\{Mam\left(\cdot\right), DConv\left(\cdot\right)\right\}$ represent the Mamba-Block and the Dilated-Block, respectively. $MEN_{out}$ represents the output of MEN branch. The CEN branch processes data in a similar way to the MEN branch, with the only difference lying in the final module. The F2E-Real and F2E-Imag modules refer to the implicit extraction of phase information by obtaining complex spectral features:
\begin{equation}
E= Mam(C_{ri})
\end{equation}
\begin{equation}
E_{r}=PReLU(LN(DConv(E)))
\end{equation}
\begin{equation}
E_{i}=PReLU(LN(DConv(E)))
\end{equation}
\begin{equation}
CEN_{out}=E_{r} \oplus E_{i}
\end{equation}
where $\oplus$ denotes the elementary sum operation. $CEN_{out}$ represents the output of CEN branch, and $E_{r}$, $E_{i}$ represent the real and imaginary parts outputted by different decoder modules, respectively.

\subsection{Mamba-Block}
\begin{figure}[t]
\centering
\includegraphics[width=1\columnwidth]{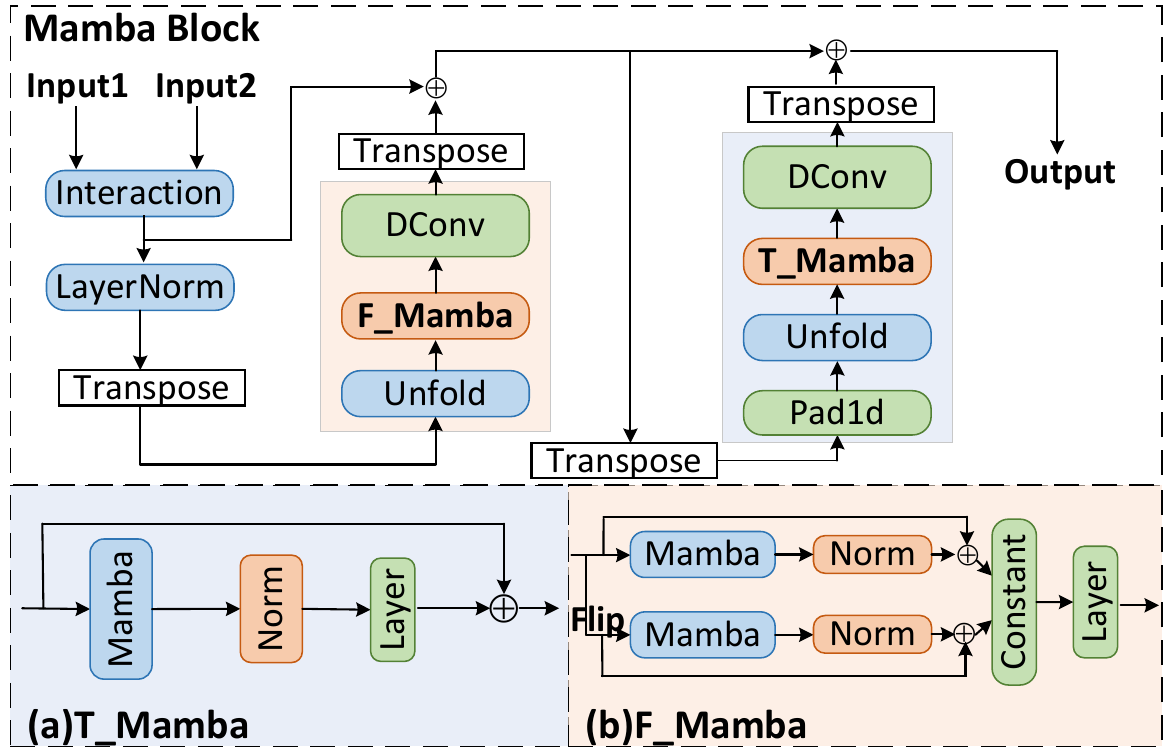} 
\caption{The Mamba-Block: It is primarily divided into temporal modeling and frequency modeling. (a) The proposed unidirectional Mamba module. (b) The proposed bidirectional Mamba module.}
\label{fig:fig3}
\end{figure}
As shown in Figure~{\ref{fig:fig3}}, the Mamba-Block module is used for sequence modeling in BSDB-Net.
The corresponding equation is as follows:
\begin{equation}
h_{n}=\mathbf{\overline{A}}h_{n-1}+\mathbf{\overline{B}}x_{n}
\end{equation}
\begin{equation}
y_{n}=\mathbf{C}h_{n}
\end{equation}
where $\mathbf{\overline{A}}$ and $\mathbf{\overline{B}}$ represents the parameters of the discretization matrix. The process of discretization converts continuous parameters $(\mathbf{\Delta},\mathbf{A},\mathbf{B})$ into discrete ones $(\mathbf{\overline{A}},\mathbf{\overline{B}})$, enabling the model to handle discrete data effectively. Mamba incorporates Selective SSMS into an H3 structure.

First, the Mamba-Block receives data from both the MEN and CEN branches enters the interaction layer for fusion. After passing through the LN layer, the dimension is transposed from $Z\in \mathbb{R}^{B\times T\times N\times F}$ to $Z\in \mathbb{R}^{(B\times T)\times N\times F}$ to better model the frequency dimension. Subsequently, the data is sent into the F\_Mamba module for sequence modeling, as illustrated in Figure 3(b). This module achieves bidirectional modeling, the forward and backward data are sent separately into the Mamba module:
\begin{equation}
X_{in}=Inter(X_{ri},X_{mag}),
F_{in}=Tran(LN(X_{in}))
\end{equation}
\begin{equation}
F_{out}=X_{in}\oplus Deconv(F\_Ma(Unfold(F_{in})))
\end{equation}
where $X_{ri}$, $X_{mag}$ represent the input of Mamba-Block in the Dual-branch. Inter represents interaction layer. F\_Ma as shown in Figure 3(b) represents sequence modeling.

The modeling for the time dimension is similar to that for the frequency dimension. However, the key difference is that unidirectional Mamba is employed for time modeling since this model operates under a causal framework. Firstly, the data will be Transpose $Z\in \mathbb{R}^{(B\times K)\times N\times T}$. Subsequently, it will go through the T-Mamba module as shown in Figure 3(a) for sequence modeling. Once the modeling is completed, the data will pass through a deconvolution layer and be Transposed into $Z\in \mathbb{R}^{B\times T\times N\times F}$:
\begin{equation}
T_{in}=Unfold(Pad(Tran(F_{out})))
\end{equation}
\begin{equation}
T_{out}=Tran(Deconv(T-Ma(T_{in})))
\end{equation}
\begin{equation}
Mamba_{out}=F_{out}\oplus T_{out}
\end{equation}
where T\_Ma as shown in Figure 3(a) represents sequence modeling, $Mamba_{out}$ represents the output of Mamba-Block. Trans represents the tensor dimension reshaping.

\subsection{Interaction Module}
We achieve information interaction through the interaction module. This module first concatenates $Input_{1}$ and $Input_{2}$ using a Cat layer, then passes them through a Conv2d layer and a LayerNorm layer, followed by generating a mask using a Sigmoid activation function. After that, it multiplies the result with $Input_{1}$ to output the final data:
\begin{equation}
Input =Cat(Input_{1},Input_{2})
\end{equation}
\begin{equation}
Mask=Sigmoid(LN(Conv2d(Input)))
\end{equation}
\begin{equation}
Output=Input_{1} + (Input_{2}\otimes Mask)
\end{equation}

\begin{table*}[t]
\raggedright
\renewcommand{\arraystretch}{0.8} 
\setlength{\tabcolsep}{5.3pt} 
\begin{tabular}{cc|c|lccc|cccc|cccc}
\toprule
\multicolumn{2}{c|}{Metrics}                              & \multirow{2}{*}{Feat.} & \multicolumn{4}{c|}{PESQ}                                                                                  & \multicolumn{4}{c|}{ESTOI(\%)}                                                                                 & \multicolumn{4}{c}{SI-SDR(dB)}                                                                                \\ \cline{1-2} \cline{4-15} 
\multicolumn{2}{c|}{SNR(dB)}                              &                        & \multicolumn{1}{c}{-5}   & 0                        & 5                        & Avg.                      & -5                        & 0                         & 5                         & Avg.                       & -5                        & 0                         & 5                         & Avg.                      \\ \midrule
\multicolumn{1}{c|}{\multirow{12}{*} { \rotatebox{90}{Set-A} }} & Noisy      & -                      & \multicolumn{1}{c}{1.54} & 1.86                     & 2.17                     & 1.85                      & 29.25                     & 43.11                     & 57.53                     & 43.30                      & -5.00                     & 0.00                      & 5.00                      & 0.00                      \\
\multicolumn{1}{c|}{}                        & ConvTasNet & Wave               & \multicolumn{1}{c}{2.11} & 2.54                     & 2.88                     & 2.52                      & 60.06                     & 73.80                     & 82.90                     & 72.25                      & 6.56                      & 10.43                     & 13.63                     & 10.21                     \\
\multicolumn{1}{c|}{}                        & DPRNN      & Wave               & \multicolumn{1}{c}{2.17} & 2.60                     & 2.96                     & 2.57                      & 61.74                     & 74.74                     & 83.53                     & 73.34                      & 6.88                      & 10.60                     & 13.82                     & 10.43                     \\
\multicolumn{1}{c|}{}                        & DDAEC      & Wave               & \multicolumn{1}{c}{2.27} & 2.79                     & 3.16                     & 2.74                      & 63.12                     & 76.65                     & 84.73                     & 74.83                      & 7.22                      & 11.23                     & 14.15                     & 10.87                     \\
\multicolumn{1}{c|}{}                        & LSTM       & Mag                     & \multicolumn{1}{c}{1.97} & 2.37                     & 2.67                     & 2.34                      & 49.33                     & 64.14                     & 74.98                     & 62.82                      & 2.49                      & 6.58                      & 9.54                      & 6.20                      \\
\multicolumn{1}{c|}{}                        & CRN        & Mag                     & \multicolumn{1}{c}{1.97} & 2.45                     & 2.79                     & 2.41                      & 50.52                     & 66.21                     & 77.24                     & 64.66                      & 2.66                      & 7.23                      & 10.79                     & 6.89                      \\
\multicolumn{1}{c|}{}                        & GCRN       & RI                     & \multicolumn{1}{c}{2.02} & 2.55                     & 2.92                     & 2.50                      & 56.44                     & 72.83                     & 82.08                     & 70.45                      & 5.36                      & 9.72                      & 12.67                     & 9.25                      \\
\multicolumn{1}{c|}{}                        & DCCRN      & RI                     & \multicolumn{1}{c}{1.90} & 2.46                     & 2.84                     & 2.40                      & 50.98                     & 68.06                     & 78.73                     & 65.92                      & 4.17                      & 8.61                      & 11.74                     & 8.17                      \\
\multicolumn{1}{c|}{}                        & FullSubNet & RI                     & \multicolumn{1}{c}{2.20} & 2.64                     & 2.97                     & 2.60                      & 50.44                     & 67.34                     & 78.88                     & 65.56                      & 4.34                      & 9.01                      & 12.81                     & 8.72                      \\
\multicolumn{1}{c|}{}                        & CTSNet     & M+RI                 & \multicolumn{1}{c}{2.32} & 2.79                     & 3.14                     & 2.75                      & 62.92                     & 76.20                     & 84.35                     & 74.49                      & 6.75                      & 10.84                     & 13.96                     & 10.52                     \\
\multicolumn{1}{c|}{}                        & GaGNet     & M+RI                 & 2.36                     & 2.85                     & 3.22                     & 2.81                      & 65.84                     & 78.13                     & 85.79                     & 76.59                      & 7.36                      & 11.23                     & 14.31                     & 10.97                     \\
\multicolumn{1}{c|}{}                        & \textbf{BSDB(ours)}   & M+RI            &
                    \multicolumn{1}{c}{\textbf{2.43}}    & \textbf{2.92}                        & \textbf{2.26}                        & \textbf{2.87}                         & \textbf{66.27}                         & \textbf{78.19}                         & \textbf{86.66}                         &\textbf{77.04}                          & \textbf{7.42}                         & \textbf{11.34}                        & \textbf{14.39}                        & \textbf{11.05}                        \\ \midrule
\multicolumn{1}{l|}{\multirow{12}{*}{\rotatebox{90}{Set-B}}} & Noisy      & -                      & 1.74                     & \multicolumn{1}{l}{2,04} & \multicolumn{1}{l}{2.41} & \multicolumn{1}{l|}{2.06} & \multicolumn{1}{l}{44.59} & \multicolumn{1}{l}{57.38} & \multicolumn{1}{l}{69.45} & \multicolumn{1}{l|}{57.14} & -5.00                     & 0.00                      & 5.00                      & 0.00                      \\
\multicolumn{1}{l|}{}                        & ConvTasNet & Wave              & 2.57                     & \multicolumn{1}{l}{2.90} & \multicolumn{1}{l}{3.21} & \multicolumn{1}{l|}{2.89} & \multicolumn{1}{l}{73.00} & \multicolumn{1}{l}{81.79} & \multicolumn{1}{l}{87.90} & \multicolumn{1}{l|}{80.90} & \multicolumn{1}{l}{10.25} & \multicolumn{1}{l}{13.18} & \multicolumn{1}{l}{16.07} & \multicolumn{1}{l}{13.17} \\
\multicolumn{1}{l|}{}                        & DPRNN      & Wave               & 2.66                     & \multicolumn{1}{l}{2.98} & \multicolumn{1}{l}{3.27} & \multicolumn{1}{l|}{2.97} & \multicolumn{1}{l}{74.23} & \multicolumn{1}{l}{82.44} & \multicolumn{1}{l}{88.32} & \multicolumn{1}{l|}{81.66} & \multicolumn{1}{l}{10.49} & \multicolumn{1}{l}{13.37} & \multicolumn{1}{l}{16.17} & \multicolumn{1}{l}{13.34} \\
\multicolumn{1}{l|}{}                        & DDAEC      & Wave               & 2.83                     & \multicolumn{1}{l}{3.17} & \multicolumn{1}{l}{3.43} & \multicolumn{1}{l|}{3.15} & \multicolumn{1}{l}{75.57} & \multicolumn{1}{l}{83.65} & \multicolumn{1}{l}{89.03} & \multicolumn{1}{l|}{82.75} & \multicolumn{1}{l}{10.91} & \multicolumn{1}{l}{13.67} & \multicolumn{1}{l}{16.16} & \multicolumn{1}{l}{13.58} \\
\multicolumn{1}{l|}{}                        & LSTM       & Mag                     & 2.37                     & \multicolumn{1}{l}{2.69} & \multicolumn{1}{l}{2.94} & \multicolumn{1}{l|}{2.67} & \multicolumn{1}{l}{64.27} & \multicolumn{1}{l}{74.51} & \multicolumn{1}{l}{81.76} & \multicolumn{1}{l|}{73.51} & \multicolumn{1}{l}{6.30}  & \multicolumn{1}{l}{9.13}  & \multicolumn{1}{l}{11.08} & \multicolumn{1}{l}{8.84}  \\
\multicolumn{1}{l|}{}                        & CRN        & Mag                     & 2.48                     & \multicolumn{1}{l}{2.83} & \multicolumn{1}{l}{3.16} & \multicolumn{1}{l|}{2.82} & \multicolumn{1}{l}{66.46} & \multicolumn{1}{l}{77.30} & \multicolumn{1}{l}{85.01} & \multicolumn{1}{l|}{76.26} & \multicolumn{1}{l}{7.04}  & \multicolumn{1}{l}{10.43} & \multicolumn{1}{l}{13.41} & \multicolumn{1}{l}{10.29} \\
\multicolumn{1}{l|}{}                        & GCRN       & RI                     & 2.55                     & \multicolumn{1}{l}{2.94} & \multicolumn{1}{l}{3.21} & \multicolumn{1}{l|}{2.90} & \multicolumn{1}{l}{70.31} & \multicolumn{1}{l}{80.82} & \multicolumn{1}{l}{86.91} & \multicolumn{1}{l|}{79.35} & \multicolumn{1}{l}{9.01}  & \multicolumn{1}{l}{12.27} & \multicolumn{1}{l}{14.69} & \multicolumn{1}{l}{11.99} \\
\multicolumn{1}{l|}{}                        & DCCRN      & RI                     & 2.47                     & \multicolumn{1}{l}{2.89} & \multicolumn{1}{l}{3.17} & \multicolumn{1}{l|}{2.84} & \multicolumn{1}{l}{66.43} & \multicolumn{1}{l}{77.92} & \multicolumn{1}{l}{84.66} & \multicolumn{1}{l|}{76.34} & \multicolumn{1}{l}{8.40}  & \multicolumn{1}{l}{11.74} & \multicolumn{1}{l}{13.82} & \multicolumn{1}{l}{11.32} \\
\multicolumn{1}{l|}{}                        & FullSubNet & RI                     & 2.66                     & \multicolumn{1}{l}{3.00} & \multicolumn{1}{l}{3.30} & \multicolumn{1}{l|}{2.99} & \multicolumn{1}{l}{67.08} & \multicolumn{1}{l}{78.20} & \multicolumn{1}{l}{85.81} & \multicolumn{1}{l|}{77.03} & \multicolumn{1}{l}{8.87}  & \multicolumn{1}{l}{12.30} & \multicolumn{1}{l}{15.60} & \multicolumn{1}{l}{12.26} \\
\multicolumn{1}{l|}{}                        & CTSNet     & M+RI                 & 2.86                     & \multicolumn{1}{l}{3.19} & \multicolumn{1}{l}{3.45} & \multicolumn{1}{l|}{3.17} & \multicolumn{1}{l}{76.60} & \multicolumn{1}{l}{84.12} & \multicolumn{1}{l}{89.28} & \multicolumn{1}{l|}{83.33} & \multicolumn{1}{l}{10.99} & \multicolumn{1}{l}{13.84} & \multicolumn{1}{l}{16.44} & \multicolumn{1}{l}{13.75} \\
\multicolumn{1}{l|}{}                        & GaGNet     & M+RI                 & \textbf{2.94}                     & \multicolumn{1}{l}{3.28} & \multicolumn{1}{l}{3.56} & \multicolumn{1}{l|}{\textbf{3.26}} & \multicolumn{1}{l}{\textbf{78.24}} & \multicolumn{1}{l}{85.30} & \multicolumn{1}{l}{90.10} & \multicolumn{1}{l|}{84.55} & \multicolumn{1}{l}{11.23} & \multicolumn{1}{l}{14.00} & \multicolumn{1}{l}{16.56} & \multicolumn{1}{l}{13.93} \\
\multicolumn{1}{l|}{}                        & \textbf{BSDB(ours)}   & M+RI                 & 2.92                        & \multicolumn{1}{l}{\textbf{3.29}}    & \multicolumn{1}{l}{\textbf{3.58}}    & \multicolumn{1}{l|}{\textbf{3.26}}    & \multicolumn{1}{l}{77.80}     & \multicolumn{1}{l}{\textbf{85.64}}     & \multicolumn{1}{l}{\textbf{90.54}}     & \multicolumn{1}{l|}{\textbf{84.66}}     & \multicolumn{1}{l}{\textbf{11.47}}     & \multicolumn{1}{l}{\textbf{14.50}}    & \multicolumn{1}{l}{\textbf{17.21}}    & \multicolumn{1}{l}{\textbf{14.39}}    \\ \bottomrule
\end{tabular}
\caption{The objective is to compare the effects of different models on PESQ, ESTOI, and SI-SDR metrics for Set-A and Set-B in an unseen speaker test set.}
\label{table1}
\end{table*}

\begin{table}
\centering
\begin{tabular}{cccc}
\hline
Modle        & Cau.             & PESQ    & MACs           \\ \hline
LSTM         & \checkmark       & 2.37       & 3.69 G/s       \\
CRN          & \checkmark       & 2.45       & 2.54 G/s       \\
GCRN         & \checkmark       & 2.55       & 2.40 G/s       \\
FullSubNet   & \checkmark       & 2.64       & 29.83 G/s      \\
CTSNet       & \checkmark       & 2.79       & 5.48 G/s       \\
ConvTasNet   & \checkmark       & 2.54       & 5.22 G/s       \\
DPRNN        & \checkmark       & 2.60       & 8.47 G/s       \\
DDAEC        & \checkmark       & 2.79       & 36.85 G/s      \\
DBTNet       & \XSolidBrush     & \textbf{3.18}       & 42.64 G/s       \\
GaGNet       & \checkmark       & 2.85       & 2.81 G/s        \\ \midrule
64-4(our)    & \checkmark       & 2.62       & \textbf{0.88 G/s}       \\
64-6(our)    & \checkmark       & 2.70       & 0.98 G/s       \\
128-4(our)   & \checkmark       & 2.78       & 1.34 G/s          \\
128-6(our)   & \checkmark       & 2.92       & 1.68 G/s          \\
256-2(our)   & \checkmark       & 2.74       & 1.87 G/s          \\
256-4(our)   & \checkmark       & 2.83       & 3.06 G/s          \\
256-6(our)   & \checkmark       & 2.95       & 4.26 G/S                \\ \hline
\end{tabular}
\caption{The aim is to compare different models and the impact of varying input dimensions and module stacking times in Mamba on PESQ and Computational Complexity. 
 (Here, ``64-4" represents an input dimension of 64 with modules stacked four times, and so forth.)}
\label{table2}
\end{table}

\section{Experimental Setup}
\subsection{Datasets}
We selected the WSJ0-SI84 and the DNS-Challenge noise dataset to create synthetic data to evaluate our model and conduct ablation experiments. Subsequently, we compared our model with others using a widely-used dataset, VoiceBank+Demand.

WSJ0-SI84+DNS-Challenge: WSJ0-SI84~\cite{wsj0} consists of 7138 clean speech samples from 83 speakers. From these, 5428 and 957 utterances from 77 speakers are randomly selected for the training and validation sets. To construct "noisy-clean" training pairs, approximately 20,000 types of noise from the DNS-Challenge~\cite{DNSChallenge} dataset's noise library are randomly selected and concatenated, resulting in a total duration of approximately 55 hours. The training set and validation set of the dataset are synthesized by the author Li~\cite{ctsnet}, and the Set-A and Set-B are consistent with the literature ~\cite{gagnet}. 

VoiceBank+Demand: VoiceBank~\cite{voicebank} consists of 30 speakers, with 28 speakers used for the training set and the remaining two speakers used for testing. The training~\cite{usevoicebank} set includes 11,572 ``noisy-clean" pairs, mixed with 10 types of noise (8 from the Demand~\cite{deman} noise database and two from artificial noise), this is a public dataset in SE, and this paper uses the same dataset as mentioned above.

\subsection{Implementation Setup}
All speech signals in the training set are sampled at 16 kHz. The speech signals are framed using a 20 ms Hann window with 50\% overlap between frames. Transforming these framed signals into the time-frequency domain involves using a 320-point FFT. Following findings from the literature~\cite{YASUOYOUXIAO1}, we use a power spectral density compression strategy, where the compression coefficient is set to 0.5, denoted as {$|X|_{0.5}$, $|S|_{0.5}$}. The Adam optimizer is utilized with parameters $\beta=0.9 $ and $\beta =0.999$. The learning rate is initialized to 5e-4, and if the validation loss does not decrease for two consecutive evaluations, the learning rate is halved.

\subsection{Baseline Models}
On the WSJ0-SI84 dataset, a total of 9 baseline methods were selected for comparison with the proposed model. ConvTasNet, DPRNN~\cite{DPRNN}, and DDAEC~\cite{ddaec} are all time-domain SE models. LSTM~\cite{LSTM}, CRN~\cite{CRN}, GCRN~\cite{GCRN}, DCCRN~\cite{dccrn}, FullSubNet~\cite{fullsubnet}, CTSNet~\cite{ctsnet}, and GaGNet~\cite{gagnet} are models in the time-frequency domain. Among them, GCRN and DCCRN were developed based on CRN. FullSubNet introduces full-band modeling and sub-band modeling. CTSNet and GaGNet are respectively parallel and serial two-stage amplitude-phase decoupling models.

On the VoiceBank+Demand dataset, a total of 13 baseline methods were selected for comparison with the proposed model. SEGAN~\cite{segan}, MMSEGAN~\cite{mmsegan}, MetricGAN~\cite{metricgan}, and SRTNET~\cite{SRTNET} are generative models of speech enhancement (SE). Wavenet~\cite{wavenet} operates as a time-domain model. PHASEN~\cite{c:5}, MHSASPK~\cite{MHSA-SPK}, DCRNN, TSTNN~\cite{tstnn}, S4NDUNet~\cite{S4ND-UNET}, FDFnet~\cite{FDFnet}, CSTnet, and GaGnet are all models in the time-frequency domain, with the latter five being multi-stage SE models. CompNet~\cite{compnet} is a model that spans both the time-domain and time-frequency domain.

\subsection{Loss Function}
Our BSDB-Net employs a dual-branch approach for speech enhancement, utilizing the ``RI+Mag" loss function to supervise the optimization of both phase and magnitude components simultaneously:
\begin{equation}
\mathcal{L}_{RI}=\left \|\tilde{S}_{r}-S_r  \right \| ^2_F + \left \|\tilde{S}_{i}-S_i  \right \| ^2_F
\end{equation}
\begin{equation}
\mathcal{L}_{Mag}=\left \| \sqrt{|\tilde{S}_{r}|^2+|\tilde{S}_{i}|^2} + \sqrt{|{S}_{r}|^2+|{S}_{i}|^2}\right \|^2_{F} 
\end{equation}
\begin{equation}
\mathcal{L}=\beta \mathcal{L}_{RI}+(1-\beta)\mathcal{L}_{Mag}
\end{equation}
where $\left \| .\right \|_{F}$ represents Frobenius norm, and $\beta$ is empirically set to 0.5.

\subsection{Evaluation Metrics}
Multiple objective metrics are adopted, including narrow-band (NB) and wide-band (WB) perceptual evaluation speech quality (PESQ)~\cite{pesq} for speech quality, short-time objective intelligibility (STOI) and its extended version ESTOI~\cite{estoi} for intelligibility, SISDR~\cite{mosSDR} for speech distortion, and MOS (CSIG, CBAK, COVL)~\cite{mosSDR} for speech quality.

\section{Results and Analysis}
\subsection{Ablation Study}
We conducted ablation experiments on the WSJ0 dataset, which cover the following three aspects: (1) Whether the dual-path structure is effective; (2) How many layers of Mamba-based modules should be stacked and how many hidden layers should be fed into Mamba to achieve the best effect; (3) Whether using Mamba as a sequence modeling model is effective.
\begin{table}
\centering
\begin{tabular}{cccc}
\toprule
Model            & PESQ & ESTOI     & SI-SDR(dB) \\ \midrule
BSDB-MEN         & 2.67     &71.25          &9.84           \\
BSDB-CEN         & 2.76     &74.32          &10.37           \\
\textbf{BSDB-DB}          & \textbf{2.92} &\textbf{78.19} &\textbf{11.34}             \\ \bottomrule
\end{tabular}
\caption{Compare the effects of enhancing magnitude spectrum only(BSDB-MEN), enhancing complex spectrum only(BSDB-CEN), and enhancing both magnitude and complex spectra in parallel(BSDB-DB) on PESQ, ESTOI, and SDR.}
\label{table3}
\end{table}

\textbf{Effect of Dual-Branch Model}: Here we are primarily investigating the effectiveness of the dual-branch structure. Initially, we removed the CEN branch to train the MEN single branch, and subsequently removed MEN while retaining the CEN branch for training. As shown in Table 3, the dual-path structure outperforms the single-path structure across all metrics. This means that the coordinated efforts of CEN and MEN can improve the quality of the target speech. CEN filters out primary noise for rough estimation, while MEN continuously supplements speech information, thereby enhancing the overall performance of the system.

\textbf{The Number of Layers and Hidden Layers of Mamba-Block}: The dimensions of the input data to Mamba have an impact on the model's performance. Additionally, stacking Mamba-Block layers can further enhance model performance but also increases complexity. As shown in Table ~{\ref{table2}, we selected Mamba input dimensions of 64, 128, and 256, and stacked modules 2, 4, and 6 times in different combinations. From our findings, when we stack sequence modeling modules and increase the number of Mamba hidden layers, performance will improve and the increase in depth has a more significant effect than the increase in breadth, using a combination like 128-6 for Mamba already achieves a balance between performance and complexity. Additionally, the current SE models are all at the complexity level of Gb/s, while our model further compresses the complexity to the level of Mb/s without a significant decrease in performance.

\textbf{Effect of Mamba-Block}: Our model aims to ensure performance while compressing model complexity, benefiting from advancements in Selective State Spaces. To demonstrate the effectiveness of Mamba-Block, we replaced it with LSTM and Transformer while keeping other factors constant, as shown in Table ~{\ref{table5}} Our results indicate that our model maintains optimal performance with significantly reduced complexity, validating the applicability and feasibility of Mamba for our task. Although the performance improvement over the Transformer is relatively minor, the significant reduction in complexity is the reason why our model opts for Mamba for sequence modeling.

\setlength{\tabcolsep}{1.5pt}
\begin{table}
\centering
\small
\begin{tabular}{c|c|ccccc}
\toprule
Modle          & Year & PESQ-WB & STOI\% & CSIG & CBAK & COVL \\ \midrule
Noisy          & -    & 1.97    & 92.1   & 3.35 & 2.44 & 2.63 \\
SEGAN          & 2017 & 2.16    & 92.5   & 3.48 & 2.94 & 2.80 \\
MMSEGAN        & 2018 & 2.53    & 93     & 3.8  & 3.12 & 3.14 \\
Wavenet        & 2018 & -       & -      & 3.62 & 3.32 & 2.98 \\
MetricGAN      & 2019 & 2.86    & -      & 3.99 & 3.18 & 3.42 \\
DCCRN          & 2020 & 2.68    & 93.7   & 3.88 & 3.18 & 3.27 \\
PHASEN         & 2020 & 2.99    & -      & 4.21 & 3.55 & 3.62 \\
MHSA-SPK       & 2020 & 2.99    & -     & 4.15 & 3.42 & 3.53 \\
TSTNN          & 2021 & 2.96    & 95     & 4.17 & 3.53 & 3.49 \\
CTS-Net        & 2022 & 2.92    & -     & 4.25 & 3.46 & 3.59 \\
GaGnet         & 2022 & 2.94    & 94.7   & 4.26 & 3.45 & 3.59 \\
SRTNET         & 2023 & 2.69    & -    & 4.12 & 3.19 & 3.39 \\
CompNet        & 2023 & 2.90    & -      & 4.16 & 3.37 & 3.53 \\
FDFNet         & 2024  & 3.05     & -     & 4.23 & 3.55 & 3.65 \\
S4DSE          & 2024 & 2.55    & -     & 3.94 & 3.00 & 3.32 \\ \midrule
\textbf{BSDBNet(256)} & 2024 & \textbf{3.11}    & \textbf{95} & \textbf{4.33} & \textbf{3.58} & \textbf{3.73} \\
\textbf{BSDBNet(128)} & 2024 & 3.07    & 94.8   & 4.32 & 3.58 & 3.71 \\ \bottomrule
\end{tabular}
\caption{Comparison was conducted with other state-of-the-art methods, including both time-domain and time-frequency domain approaches. ``-" indicates where results were not provided in the original text.}
\label{table4}
\end{table}

\subsection{Model Complexity Comparison}
As shown in Table ~{\ref{table2}}, we evaluated the complexity of our proposed model and other baseline models on the WSJ0-SI84 dataset. It is worth noting that all input samples were set to one second of audio, ensuring fairness in our experiments. The model selected from the previous ablation experiments has an average computational complexity about 8 times lower than the baseline and has good performance. BSDB-Net exhibits lower complexity along with excellent performance. In terms of performance, our model is only slightly behind DBTNet. The reasons for this can be analyzed as follows: First, our model is a causal model while DBTNet is a non-causal model. Second, DBTNet has been continuously increasing its model complexity to achieve SOTA performance. As can be seen from Table ~{\ref{table2}}, its model complexity is about 20 times that of our model, yet the performance has not been significantly improved. When we further compress the complexity to the level of M/s, proving the effectiveness of our network structure.

\subsection{Comparisons with Baselines on WSJ0-SI84 Corpus}
As shown in Table ~{\ref{table1}}, the objective metrics results for the proposed method and baseline models on the WSJ0-SI84 dataset include PESQ, ESTOI, and SI-SDR. From the table, we can draw the following conclusions:
Firstly, it is evident that models based on complex spectra generally outperform those based solely on magnitude spectra. For instance, models like GCRN and DCCRN consistently outperform CRN and LSTM across all metrics. This indicates that incorporating phase information in addition to amplitude recovery can significantly enhance both speech quality and intelligibility. Secondly, multi-stage models that simultaneously consider magnitude and complex spectra outperform models that focus solely on a single spectrum type or operate in the time domain. Models like CSTNet and GaGNet show superior performance compared to others, suggesting that parallel optimization of amplitude and complex spectra can effectively leverage phase information to generate higher-quality speech. Lastly, our model achieves state-of-the-art (SOTA) results across all metrics. The reasons for this can be analyzed as follows: BSDB-NET employs a parallel two-stage model, which is better at decoupling amplitude and phase information to address compensation effects compared to other models; A sequence modeling module based on Mamba is constructed, demonstrating good performance at a linear complexity.
\subsection{Comparisons with Baselines on VoiceBank + Demand}
In addition to the WSJ0-SI84 corpus, we also conducted experiments on another public benchmark, VoiceBank + Demand. BSDB-Net was compared with other baselines, and our model achieved superior results across all metrics. As evident from Table ~{\ref{table4}}, the baseline models achieved average improvements of 0.32, 1.5\%, 0.31, 0.17, and 0.35 in PESQ, STOI, CSIG, CBAK, and COVL, respectively. This indicates that the methodologies proposed by our model adeptly address the issues presented in the background. The main purpose of experimenting on this public dataset is to fairly demonstrate that our model can maintain competitive performance while significantly reducing complexity, thereby better proving that the network has indeed effectively solved the proposed problem.
\begin{table}
\centering
\begin{tabular}{cccc}
\toprule
Model            & PESQ & MACs     & Parameters \\ \midrule
Ours-LSTM        & 2.82 & 11.05G/s & 14.71M     \\
Ours-Transformer & 2.90 & 28.06G/s & 14.06M          \\
\textbf{Ours-Mamba(ours)}       & \textbf{2.92} & \textbf{1.68G/s}  & \textbf{9.78M}       \\ \bottomrule
\end{tabular}
\caption{Compare the impact of replacing the sequence modeling module with LSTM, Transformer, and Mamba on PESQ, Computational Complexity, and Parameters.}
\label{table5}
\end{table}
\section{Conclusion}
In this paper, we propose a Band-Split Dual-branch Network based on Selective State Spaces. This network reduces complexity while estimating speech spectra through a complementary mechanism. Specifically, we divide the network into a magnitude enhancement net (MEN) and a complex spectral enhancement net (CEN), which jointly filter noise components while continuously refining and supplementing speech spectrum information. To further reduce model complexity, we introduce a band-splitting strategy in each branch. Additionally, we incorporate Mamba for sequence modeling, ensuring model performance while reducing complexity. In future work, we intend to further enhance model performance and decrease the computational complexity. Besides, we intend to apply the proposed method to more tasks, like multi-channel speech enhancement, dereverberation, and target extraction. Mamba is still far from mature in deployment, which renders it necessary for further optimization.
\section{Acknowledgments}
This work is supported by the {STI 2030—Major Projects (No. 2021ZD0201500)}, the National Natural Science Foundation of China (NSFC) (No.62201002, 6247077204), Excellent Youth Foundation of Anhui Scientific Committee (No. 2408085Y034), Distinguished Youth Foundation of Anhui Scientific Committee (No. 2208085J05), Special Fund for Key Program of Science and Technology of Anhui Province (No. 202203a07020008), Cloud Ginger XR-1.
\bibliography{aaai25.bbl}

\end{document}